\documentclass[prd,aps,eqsecnum]{revtex4}
\usepackage{graphicx}
\newcommand{\be}{\begin{equation}}
\newcommand{\ee}{\end{equation}}
\newcommand{\Be}{\begin{eqnarray}}
\newcommand{\Ee}{\end{eqnarray}}
\newcommand{\f}{\frac}
\newcommand{\pa}{\partial}
\begin{document}
\title{Gravitational perturbation of traversable wormhole}
\author{Sung-Won Kim\footnote{E-mail address: sungwon@mm.ewha.ac.kr}}
\affiliation{Department of Science Education, Ewha Women's
University, Seoul 120-750, Korea}

\date{\today}

\begin{abstract}

In this paper, we study the perturbation problem of the scalar,
electromagnetic, and gravitational waves under the traversable
Lorentzian wormhole geometry. The unified form of the potential
for the Schr\"odinger type one-dimensional wave equation is found.

\end{abstract}

\maketitle


\section{Introduction}

The wormhole has the structure which is given by two
asymptotically flat regions and a bridge connecting two
regions\cite{MT88}. For the Lorentzian wormhole to be traversable,
it requires exotic matter which violates the known energy
conditions. To find the reasonable models, there had been studying
on the generalized models of the wormhole with other matters
and/or in various geometries. Among the models, the matter or wave
in the wormhole geometry and its effect such as radiation are very
interesting to us. The scalar field could be considered in the
wormhole geometry as the primary and auxiliary effects\cite{K00}.
Recently, the solution for the electrically charged case was also
found \cite{KL01}.

Scalar wave solutions in the wormhole geometry\cite{KSB94,KMMS95}
was in special wormhole model only and the transmission and
reflection coefficients were found.  The electromagnetic wave in
wormhole geometry is recently discussed\cite{BH00} along the
method of scalar field case. These wave equations in wormhole
geometry draws attention to the research on radiation and wave.

Also there was a suggestion that the wormhole would be one of the
candidates of the gamma ray bursts\cite{TRA98}. With such
suggestions, we can also suggest the wormhole as one of the
candidates of the gravitational wave sources. If the gravitational
wave detections are achieved in future, the identification of the
wormhole might be followed by the unique waveforms from the
perturbed exotic matter consisting of wormhole.

For the gravitational radiation in any forms, the scattering
problem to calculate the cross section in more generalized models
of wormhole should be considered. Thus the study of scalar,
electromagnetic, and gravitational waves under wormhole geometry
is necessary to the research on the gravitational radiation.

In this paper, we found the general form of the gravitational
perturbation of the traversable wormhole, which will be a key to
extend the wormhole physics into the problems similar to those
relating to gravitational wave of black holes. The main idea and
resultant equation is similar to Regge-Wheeler equation\cite{RW57}
for black hole perturbation. Here we adopt the geometrical unit,
{\it i.e.}, $G=c=\hbar=1$.

\section{Scalar perturbation}

The spacetime metric for static uncharged wormhole is given as \be
ds^2 = -e^{2\Lambda(r)}dt^2 + \f{dr^2}{1-b(r)/r} + r^2
(d\theta^2+\sin^2\theta d\phi^2), \label{eq:mtwormhole} \ee where
$\Lambda(r)$ is the lapse function and $b(r)$ is the wormhole
shape function. They are assumed to be dependent on $r$ only for
static case.

The wave equation of the minimally coupled massless scalar field
is given by \be \nabla^\mu\nabla_\mu\Phi =\f{1}{\sqrt{-g}}
\partial_\mu
(\sqrt{-g} g^{\mu\nu}\partial_\nu \Phi ) = 0. \ee In spherically
symmetric space-time, the scalar field can be separated by
variables, \be \Phi_{lm} = Y_{lm}(\theta, \phi)\f{u_l(r,t)}{r},
\label{eq:def1} \ee where $Y_{lm}(\theta, \phi)$ is the spherical
harmonics and $l$ is the quantum angular momentum.

If $l=0$ and the scalar field $\Phi(r)$ depends on $r$ only, the
wave equation simply becomes the following relation\cite{KK98}:
\be e^\Lambda \sqrt{1-\f{b}{r}} r^2 \f{\partial}{\partial r}\Phi =
A = \mbox{const.} \ee In this relation, the back reaction of the
scalar wave on the wormhole geometry is neglected. Thus the static
scalar wave without propagation is easily found as the integral
form of \be \Phi = A \int e^{-\Lambda} r^{-2} \left( 1- \f{b}{r}
\right)^{-1/2} dr. \ee The scalar wave solution was already given
to us for the special case of wormhole in Ref.~\cite{KL01,KK98}.

More generally, if the scalar field $\Phi$ depends on $r$ and $t$,
the wave equation after the separation of variables
$(\theta,\phi)$ becomes \be - \ddot{u}_l + \f{\partial^2
u_l}{\partial r_*^2 } = V_l \,\,u_l, \ee where the potential is
\Be V_l(r)&=&\f{L^2}{r^2}e^{2\Lambda} + \f{1}{r}e^\Lambda
\sqrt{1-\f{b}{r}}\f{\partial}{\partial r}\left(e^\Lambda
\sqrt{1-\f{b}{r}}\right) \nonumber \\
&=& e^{2\Lambda}\left[ \f{l(l+1)}{r^2} - \f{b'r-b}{2r^3} +
\f{1}{r}\left(1-\f{b}{r}\right)\Lambda' \right]\nonumber\\ \Ee and
the proper distance $r_*$ has the following relation to $r$: \be
\f{\partial}{\partial r_*} = e^\Lambda r^2 \sqrt{1-\f{b}{r}}
\f{\partial}{\partial r}. \ee Here, $L^2=l(l+1)$ is the square of
the angular momentum.

The properties of the potential are determined by the shape of it,
if only the explicit forms of $\Lambda$ and $b$ are given. If the
time dependence of the wave is harmonic as $u_l(r,t) =
\hat{u}_l(r,\omega)e^{-i\omega t} $, the equation becomes \be
\left( \f{d^2}{dr^2_*} + \omega^2 - V_l(r)
\right)\hat{u}_l(r,\omega) = 0. \ee It is just the Schr\"odinger
equation with energy $\omega^2$ and potential $V_l(r)$. When
$e^{2\Lambda}$ is finite, $ V_l $ approaches zero as $r
\rightarrow \infty $, which means that the solution has the form
of the plane wave $ \hat{u}_l \sim e^{\pm i \omega r_*}$
asymptotically. The result shows that if a scalar wave passes
through the wormhole the solution is changed from $e^{\pm i \omega
r}$ into $e^{\pm i \omega r_*}$, which means that the potential
affects the wave and experience the scattering.

As the simplest example for this problem, we consider the special
case $ (\Lambda = 0, b = b_0^2/r)$ as usual, the potential should
be in terms of $r$ or $r_*$ as \be V_l = \f{l(l+1)}{r^2} +
\f{b_0^2}{r^4} ~\stackrel{\mbox{or}}{=}~ \f{l(l+1)}{r_*^2+b_0^2} +
\f{b_0^2}{(r_*^2+b_0^2)^2}, \label{eq:pot} \ee where the proper
distance $r_*$ is given by \be r_* = \int
\f{1}{\sqrt{1-b_0^2/r^2}}dr = \sqrt{r^2-b_0^2}. \label{eq:prop1}
\ee There is the hyperbolic relation between $r_*$ and $r$ which
is plotted in Fig.~1. The potentials are depicted in Fig.~2. The
potential has the maximum value as \be V_l(r_*)|_{\rm max} =
V_l(0) = \f{l(l+1)+1}{b_0^2}. \ee

\begin{figure}
\begin{center}
\includegraphics[width=10cm]{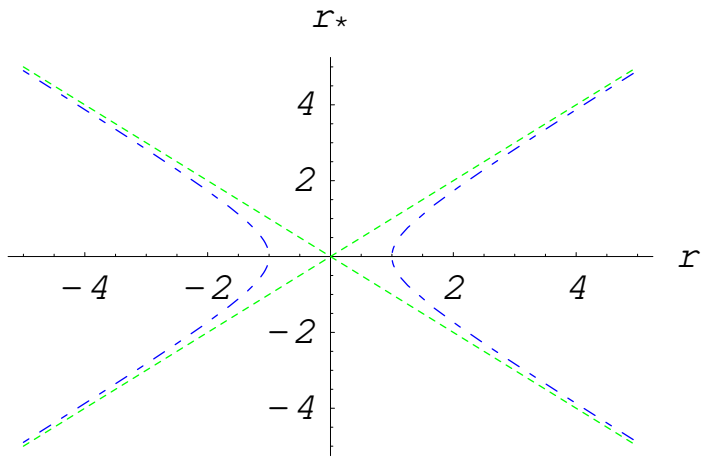}
\caption{Plot of the proper distance $r_*$ versus $r$. Here we set
$b_0=1$. The dotted line is the asymptotic line to the hyperbolic
relation, the dashed line.}
\end{center}
\end{figure}

\begin{figure}
\begin{center}
\includegraphics[width=10cm]{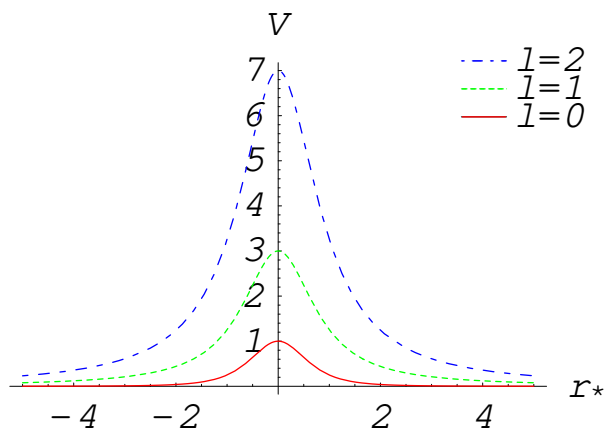}
\caption{Plot of the potentials of the scalar wave under the
specified wormhole $b=b_0^2/r$ in terms of $r_*$ for $l=1,2,3$.
Here we set $b_0=1$. To have a positive potential, $l \ge 0$. }
\end{center}
\end{figure}

\section{Electromagnetic Wave}

We just followed the result of Bergliaffa and Hibberd\cite{BH00}.
They used the electromagnetic wave under wormhole geometry of the
Morris-Thorne type wormhole like our model. Maxwell's equation in
a gravitational field is \be {H^{\mu\nu}}_{;\nu} = 4\pi I^\mu,
\quad H_{\mu\nu,\sigma}+H_{\nu\sigma,\mu}+H_{\sigma\mu,\nu} = 0,
\ee where \be H^{\mu\nu} \equiv \sqrt{-g} F^{\mu\nu}, \quad
I^{\mu} \equiv \sqrt{-g} J^{\mu}\ee and the electromagnetic field
strength tensors are defined by \be F_{\mu\nu} \rightarrow
(\vec{E}, \vec{B}), \quad H^{\mu\nu} \rightarrow (-\vec{D},
\vec{H}). \ee Defining the vectors \be \vec{F}^\pm \equiv \vec{E}
\pm i\vec{H}, \quad \vec{S}^\pm \equiv \vec{D} \pm i\vec{B}, \ee
the Einstein-Maxwell equations are \be \vec{\nabla} \wedge
\vec{F}^\pm = \pm i \f{\pa\vec{S}^\pm}{\pa t} = \pm in
\f{\pa\vec{F}^\pm}{\pa t}, \ee \be \vec{\nabla}\cdot\vec{S}^\pm =
0, \ee where $n$ is the refraction index \be \varepsilon_{ik} =
\mu_{ik} = -\sqrt{-g}\f{g_{ik}}{g_{00}} \equiv n\delta_{ik} \ee
for a medium characterized by diagonal electric and magnetic
permeabilities.

The Morris-Thorne type wormhole metric, Eq.~(\ref{eq:mtwormhole})
can be rewritten as \Be ds^2 &=& -e^{2\Lambda(r)}dt^2 + \left(
1 - \f{b(r)}{r}\right)^{-1}dr^2 + r^2 d\Omega^2 \nonumber \\
&=& -e^{2\Lambda(r)}dt^2 + A^2(\rho) (d\rho^2+\rho^2d\Omega^2),\Ee
where $A(\rho)$ is defined by \be n(\rho) =
\f{A(\rho)}{e^{\Lambda(\rho)}} \ee

Here we consider the the special case ($\Lambda(\rho)=0, b(r) =
\f{b_0^2}{r}$) like the scalar wave case. The Herz vector is
decomposed into \be \vec{F}^\pm(\rho,t) = \sum_{J,M}
\vec{F}^\pm_{JM}(\rho,t) \ee with the generalized spherical
harmonics  $\vec{Y}^{(\lambda)}_{JM}(\hat{\rho})$ \be
\vec{F}^\pm_{JM}(\rho,t) =
\sum_{\lambda=e,m,o}F^\pm_{JM}(\rho,\omega)\vec{Y}^{(\lambda)}_{JM}(\hat{\rho})e^{-i\omega
t}. \ee

The Maxwell equation becomes \be -\f{d}{d\rho}(\rho
F^{\pm(m)}_{JM}) = \pm n \omega \rho F^{\pm(e)}_{JM}, \ee \be
\f{d}{d\rho}(\rho F^{\pm(e)}_{JM}) - \sqrt{J(J+1)}F^{\pm(o)}_{JM}
= \pm n \omega F^{\pm(m)}_{JM}, \ee and \be -\f{1}{\rho}
\sqrt{J(J+1)} F^{\pm(m)}_{JM} = \pm n \omega F^{\pm(o)}_{JM}. \ee

 Let the new coordinate $x$ be \be \f{dx}{d\rho} = n(\rho), \quad x = \pm
\sqrt{r^2-b_0^2} \ee and introduce the function  \be
\chi^{\pm(\lambda)}_{JM}(x,\omega)=\rho(x)F^{\pm(\lambda)}_{JM}[\rho(x),\omega].
\ee Here $x$ plays the role of the proper distance $r_*$. The wave
equation is finally \be \f{d^2\chi^{\pm(m)}_{JM}}{dz^2} +
[\omega^2b_0^2-U_J(z)]\chi^{\pm(m)}_{JM} = 0, \ee where $z=x/b_0$
and the potential is \be U_J(z) = 4J(J+1)\left[
\f{z+\sqrt{1+z^2}}{(z+\sqrt{1+z^2})^2+1} \right]^2. \ee The
potential in our context becomes \be V(r) = \f{l(l+1)}{r^2} \ee
The potentials are depicted in Fig.~3. Here $l \ge 1$ for the
positive potential.

\begin{figure}
\begin{center}
\includegraphics[width=10cm]{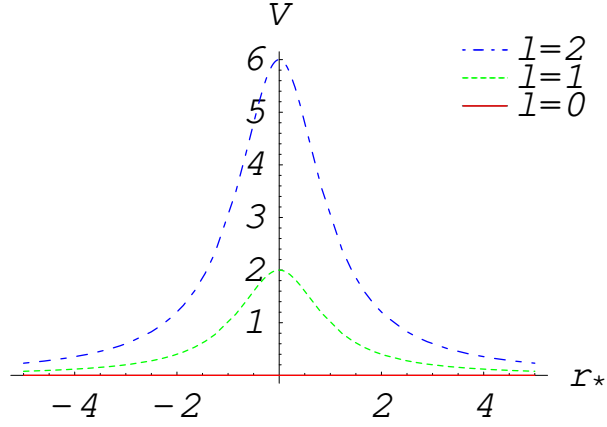}
\caption{Plot of the potentials of the electromagnetic wave under
the specified wormhole $b=b_0^2/r$ for $l=1,2,3$. Here we set
$b_0=1$. To have a positive potential, $l \ge 1$.}
\end{center}
\end{figure}

\section{Gravitational Perturbation}

We follow the conventions of Chandrasekhar in Ref.~\cite{C83}
where the gravitational perturbations are derived.  Start from the
axially symmetric spacetime which is given by \be ds^2 = -
e^{2\nu}dt^2 + e^{2\psi}(d\phi - q_1dt - q_2dr - q_3d\theta )^2 +
e^{2\mu_2}dr^2 + e^{2\mu_3}d\theta^2. \ee For unperturbed case,
the wormhole spacetime is \be e^{2\nu} = e^{2\Lambda}, \quad
e^{-2\mu_2}=\left( 1 - \f{b}{r} \right) = \f{\Delta}{r^2}, \quad
\Delta = r^2 - br, \quad e^{\mu_3} = r, \quad e^\psi = r
\sin\theta \ee and \be q_1=q_2=q_3=0. \ee

Axial perturbations are characterized by the nonvanishing of small
$q_1$, $q_2$, and $q_3$. When there are linear perturbations
$\delta\nu, \delta\psi, \delta\mu_2, \delta\mu_3$, then there are
polar perturbations with even parity which will not be considered
here. From Einstein's equation \be ( e^{3\psi+\nu-\mu_2-\mu_3}
Q_{23} )_{,3} = - e^{3\psi-\nu-\mu_2+\mu_3} Q_{02,0} \ee where
$x^2 = r, x^3 = \theta$ and $Q_{AB}=q_{A,B}-q_{B,A}, Q_{A0} =
q_{A,0}-q_{1,A}.$ This becomes \be
\f{e^\Lambda}{\sqrt{\Delta}}\f{1}{r^3\sin^3\theta}\f{\pa
Q}{\pa\theta} = - (q_{1,2} - q_{2,0})_{,0} \ee where $Q$ is \be
Q(t,r,\theta) = \Delta Q_{23}\sin^3\theta = \Delta
(q_{2,3}-q_{3,2})\sin^3\theta. \ee

 Another equation \be ( e^{3\psi+\nu-\mu_2-\mu_3} Q_{23} )_{,2} =
e^{3\psi-\nu+\mu_2-\mu_3} Q_{03,0} \ee This becomes \be
\f{e^\Lambda\sqrt{\Delta}}{r^3\sin^3\theta}\f{\pa Q}{\pa\theta} =
(q_{1,3} - q_{3,0})_{,0} \ee

If the time dependence is $e^{i\omega t}$, then \Be
\f{e^\Lambda}{\sqrt{\Delta}}\f{1}{r^3\sin^3\theta}\f{\pa
Q}{\pa\theta} &=& - i\omega q_{1,2} - \omega^2 q_2 \\
\f{e^\Lambda\sqrt{\Delta}}{r^3\sin^3\theta}\f{\pa Q}{\pa\theta}
&=& + i \omega q_{1,3} + \omega^2 q_3. \Ee

Let $Q(r,\theta) = Q(r)C^{-3/2}_{l+2}(\theta)$, where Gegenbauer
function $C^\nu_n (\theta)$ satisfy the differential equation \be
\left[ \f{d}{d\theta}\sin^{2\nu}\theta \f{d}{d\theta} +
n(n+2\nu)\sin^{2\nu}\theta \right] C^\nu_n (\theta) = 0. \ee Then
\be re^\Lambda \sqrt{\Delta} \f{d}{dr} \left( \f{e^\Lambda
\sqrt{\Delta}}{r^3} \f{dQ}{dr} \right) - \mu^2
\f{e^{2\Lambda}}{r^2}Q + \omega^2Q = 0,  \ee where
$\mu^2=(l-1)(l+2)$. If $Q=rZ$ and $\f{d}{dr_*}=e^\Lambda
\sqrt{\Delta}\f{1}{r}\f{d}{dr} $, \be \left( \f{d^2}{dr_*^2} +
\omega^2 - V(r) \right) Z = 0, \ee where the potential is \be V(r)
= e^{2\Lambda}\f{1}{r^3} \left[ \mu^2 r + \f{3\Delta}{r} - \Delta
\Lambda' - \f{1}{2} \Delta' \right]\ee or \be V(r) = e^{2\Lambda}
\left[ \f{l(l+1)}{r^2} + \f{b'r-5b}{2r^3} - \f{1}{r}\left( 1 -
\f{b}{r} \right) \Lambda' \right]\ee in terms of $b$ and
$\Lambda$. The first term is the same as the former two cases, but
the signs and coefficients of the second and third terms are
different from the scalar and electromagnetic wave cases.

For the simplest special case ($\Lambda=0, b=b_0^2/r$) like the
former cases, the potential is \be V(r) = \f{l(l+1)}{r^2} -
\f{3b_0^2}{r^4}, \ee whose shapes are shown in Fig.~4. By
comparing with scalar and electromagnetic cases, the unified
general formula is \be V(r) = \f{l(l+1)}{r^2} +
\f{(1-s^2)b_0^2}{r^4}, \ee where $s=0,1,2$ is spin, or \be V(r) =
\f{l(l+1)}{r^2} + \f{\lambda b_0^2}{r^4}, \ee where $\lambda =
(1-s^2) = 1,0,-3$ for scalar, electromagnetic, and gravitational
perturbations, respectively. This unified form is similar to the
black hole case. The condition of the positive potential is $l \ge
s$.

\begin{figure}
\begin{center}
\includegraphics[width=10cm]{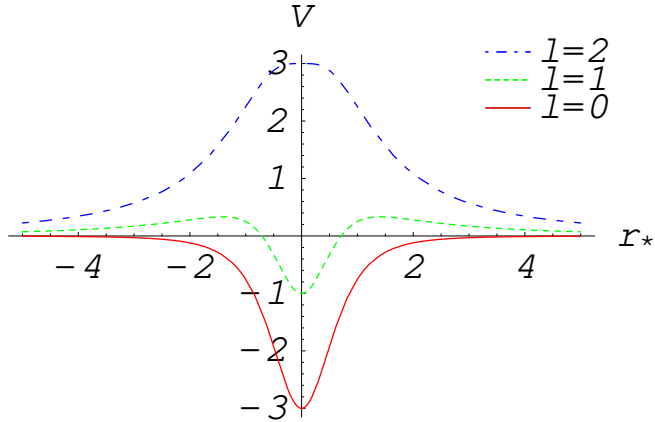}
\caption{Plot of the potentials of the gravitational wave under
the specified wormhole $b=b_0^2/r$ for $l=1,2,3$. Here we set
$b_0=1$. To have a positive potential, $l\ge 2$. }
\end{center}
\end{figure}

\section{Discussion}

We found the Regge-Wheeler type equation for gravitational
perturbation. This unified form will give us new ideas and
insights in the areas of wormhole physics and gravitational wave.
In this paper we only consider the axially perturbation for
simplicity. For further problems, Zerilli\cite{Z70} type equation
should be considered in order to see the exotic matter
perturbation, and checked whether the potential form is similar to
that of our Regge-Wheeler type potential like the black hole case
or not.

\acknowledgments

This work was supported by grant No. R01-2000-000-00015-0 from the
Korea Science and Engineering Foundation.


\begin{thebibliography}{99}

\bibitem{MT88} M. S. Morris and K. S. Thorne, Am. J. Phys. {\bf 56}, 395 (1988).


\bibitem{K00} S.-W. Kim, Gravitation \& Cosmology, {\bf 6}, 337 (2000).

\bibitem{KL01} S.-W. Kim and H. Lee, Phys. Rev. D {\bf 63},  063014 (2001).

\bibitem{KSB94} S. Kar, D. Sahdev, and B. Bhawal, Phy. Rev. D
{\bf 49}, 853 (1994).

\bibitem{KMMS95} S. Kar, S. N. Minwalla, D. Mishra, and D. Sahdev,
Phy. Rev. D {\bf 51}, 1632 (1995).

\bibitem{BH00} S. E.  P. Bergliaffa and  K. E.  Hibberd, Phy. Rev.  D {\bf 62},  044045
(2000).

\bibitem{TRA98} D. Torres,  G. Romero,  and L.  Anchordoqui, Phy.  Rev. D  {\bf 58},
123001 (1998).

\bibitem{RW57} T. Regge and J. A. Wheeler, Phys. Rev., {\bf 108},
1063 (1957).

\bibitem{KK98} S.-W. Kim and S. P. Kim, Phys. Rev. D {\bf 58}, 087703 (1998).

\bibitem{C83} S. Chandrasekhar, {\it Mathematical Theory of Black Holes} (Clarendon Press, Oxford, 1988).

\bibitem{Z70} F. J. Zerilli, Phys. Rev. D {\bf 1}, 2141 (1970).

\end{thebibliography}
\end{document}